\documentclass[]{aa}
\usepackage{natbib}         % pour A&A
\usepackage{graphicx}
\usepackage{longtable}
\usepackage{xcolor}
\usepackage[switch]{lineno} % switch=Put the line into the inner margin
\bibpunct{(}{)}{;}{a}{}{,} % to follow the A&A style

\newcommand{\ind}[1]{_{\mathrm{#1}}}
\newcommand{\diff}{\mathrm{d}}

\newcommand\Dnu{\Delta\nu}
\newcommand\Dnup{\Delta\nu (\np)}
\newcommand\Tg{\Delta\Pi_1}

\newcommand\dnurot{\delta\nu\ind{rot}}
\newcommand\dnurotmax{\delta\nu\ind{rot,core}}
\newcommand\dnurotcore{\delta\nu\ind{rot,core}}
\newcommand\dnurotenv{\delta\nu\ind{rot,env}}

\newcommand\nup{\nu\ind{p}}\newcommand\nug{\nu\ind{g}}

\newcommand\numax{\nu\ind{max}}
\newcommand\nmax{n\ind{max}}
\newcommand\xrot{x\ind{rot}}
\newcommand\DP{\Delta P}
\newcommand\Pm{\tau}
\newcommand\DPm{\Delta{\Pm_m}}\newcommand\DPmm{\Delta{\Pm_{m=-1}}}\newcommand\DPmp{\Delta{\Pm_{m=1}}}
\newcommand\thetap{\theta\ind{p}}\newcommand\thetag{\theta\ind{g}}
\newcommand\np{{n\ind{p}}}\renewcommand\ng{{n\ind{g}}}

\newcommand\nm{{n\ind{m}}}
\newcommand\epsp{\varepsilon\ind{p}}
\newcommand\epsg{\varepsilon\ind{g}}
\newcommand\nmix{\mathcal{N}}
\newcommand\zetag{\zeta\ind{g}}
\newcommand{\BV}{Brunt-V\"ais\"al\"a}
\newcommand{\NBV}{N\ind{BV}}
\newcommand{\glitchg}{g}
\newcommand{\periodeglitch}{{\mathcal{G}}}
\newcommand{\periodglitch}{{\mathcal{G}}^\star}
\newcommand\Kepler{Kepler}

\begin{document}

\title{Period spacings in red giants}
\subtitle{I. Disentangling  rotation and revealing core structure
discontinuities}
\titlerunning{Mixed modes}
\author{%
 B. Mosser\inst{1},
 M. Vrard\inst{1},
 K. Belkacem\inst{1},
 S. Deheuvels\inst{2,3},
 M.J. Goupil\inst{1}
}

\institute{
 LESIA, Observatoire de Paris, PSL Research University, CNRS, Universit\'e Pierre et Marie Curie,
 Universit\'e Paris Diderot,  92195 Meudon, France; \texttt{benoit.mosser@obspm.fr}
 \and
 Universit\'e de Toulouse; UPS-OMP; IRAP; Toulouse, France
 \and
 CNRS; IRAP; 14, avenue Edouard Belin, F-31400 Toulouse, France
 }

%\date{Submitted to A\&A}

\abstract{Asteroseismology allows us to probe the physical
conditions inside the core of red giant stars. This relies on the
properties of the global oscillations with a mixed character that
are highly sensitive to the physical properties of the core.
However, overlapping rotational splittings and mixed-mode spacings
result in complex structures in the mixed-mode pattern, which
severely complicates its identification and the measurement of the
asymptotic period spacing.}
{This work aims at disentangling the rotational splittings from
the mixed-mode spacings, in order to open the way to a fully
automated analysis of large data sets.}
{An analytical development of the mixed-mode asymptotic expansion
is used to derive the period spacing between two consecutive mixed
modes. The \'echelle diagrams constructed with the appropriately
stretched periods are used to exhibit the structure of the gravity
modes and of the rotational splittings.}
{We propose a new view on the mixed-mode oscillation pattern based
on corrected periods, called stretched periods, that mimic the
evenly spaced gravity-mode pattern. This provides a direct
understanding of all oscillation components, even in the case of
rapid rotation. The measurement of the asymptotic period spacing
and the signature of the structural glitches on mixed modes are
then made easy.}
{%
This work opens the possibility to derive all seismic global
parameters in an automated way, including the identification of
the different rotational multiplets and the measurement of the
rotational splitting, even when this splitting is significantly
larger than the period spacing. Revealing buoyancy glitches
provides a detailed view on the radiative core.}


\maketitle

%\voffset = -1.2cm
%________________________________________________________________
\section{Introduction}

Asteroseismic observations obtained with CoRoT and \emph{Kepler}
have delivered munificent information on the stellar interior
structure \citep[e.g.,][]{2008Sci...322..558M,2011Sci...332..213C}
especially for red giants
\citep[e.g.,][]{2009Natur.459..398D,2010ApJ...713L.176B}.
Independent of modelling, global seismic properties provide
relevant estimates of the stellar masses and radii
\citep[e.g.,][]{2010A&A...522A...1K,2010A&A...517A..22M}. The
dipole modes, with a mixed character, probe the core and test the
evolutionary status of the stars
\citep{2011Natur.471..608B,2011A&A...532A..86M}. They also provide
the measurement of the asymptotic period spacing
\citep{2012A&A...540A.143M} which is directly related to the core
mass \citep{2013ApJ...766..118M}. Their observation also gives
access to the differential-rotation profile in red giants
\citep{2012Natur.481...55B}. The measurement of the mean core
rotation for about 300 stars analyzed by
\cite{2012A&A...548A..10M}, extended toward a few subgiants
\citep{2014A&A...564A..27D}, indicates that angular momentum must
be efficiently transferred from the stellar core to the envelope
\citep[e.g.,][]{2013A&A...549A..74M}. At this stage, explaining
the spinning-down remains difficult, but a recent work by
\cite{2015A&A...579A..30B,2015A&A...579A..31B} shows that mixed
modes likely participate to the angular-momentum transfer and
induce the slowing down of the core rotation.

All these results illustrate the capability of the mixed-mode
oscillations to probe the innermost radiative core and deliver
unique information on the physical conditions deep inside the
star. This emphasizes the need of new observational constraints,
on data sets as large as possible. As more than 13\,000 red giant
oscillation spectra observed by \Kepler\ show solar-like
oscillations \citep{2013ApJ...765L..41S}, rotation could be
measured in a much larger data set than previously done
\citep{2012A&A...548A..10M}. This would enable us to test in
detail how the mean core rotation and the angular momentum
transfer vary with stellar evolution.

Up to now, precise rotational splittings were mostly manually
determined. Their automated measurement is possible only when the
rotational frequency is much smaller than the mixed-mode spacing.
This condition is most often met for clump stars, but barely on
the red giant branch (RGB). We seek a rapid, effective, and
automated method for measuring them in thousands of stars. This
first requires a way to disentangle rotational splittings and
period spacings, then an automated measurement of the asymptotic
period spacing.  The method recently proposed by
\cite{2015MNRAS.447.1935D} provides a partial answer for measuring
the period spacing, applicable only to non-rotating stars or stars
observed pole-on.

As is well known, rapid interior structure variation affect the
sound-speed and \BV\ frequency profiles, so that they modify the
regular arrangement of both pure pressure and pure gravity modes,
hence of the mixed-mode pattern. Such sound-speed glitches are
clearly identified in red giant oscillations
\citep{2010A&A...520L...6M,2015A&A...579A..83C,2015A&A...579A..84V};
they arise from the helium second ionization zone
\citep[e.g.,][]{2014MNRAS.440.1828B}. Their effect on the radial
oscillation modes remains limited to minor frequency changes
compared to the large separation, but is enough to induce a clear
difference between stars on the RGB or in the red clump. This
difference was observationally shown by \cite{2012A&A...541A..51K}
and theoretically studied by \cite{2014MNRAS.445.3685C}. A
systematic survey by \cite{2015A&A...579A..84V} firmly assesses
this difference and shows that it is created by acoustic glitches
in the external envelope, related to the second ionization of
helium.

Buoyancy glitches are also expected in red giants, as a
consequence of an interior structure gradient due to a local
phenomenon related to nuclear burning or mixing
\citep{2015ApJ...805..127C}. Up to now, they were not observed.
However, as suggested by \cite{2015ApJ...805..127C}, they may
explain cases where the identification of the mixed-mode pattern
is difficult or not possible.

In this work, with a new expression of the second-order asymptotic
expansion for dipole mixed modes, we analyze the structure of
their oscillation pattern including rotation, sound-speed
glitches, or buoyancy glitches. In Section~\ref{egalite}, we show
that the period spacing and the rotational splitting are
constructed on similar patterns: both are bumped near the pressure
dominated modes. The rotational pattern was formerly derived by
\cite{2015A&A...580A..96D}. Consequences of this new result are
discussed in Section \ref{applications}, where we show how the
structure of the pure gravity-mode pattern can be revealed from
the mixed-mode pattern. Section \ref{disentangling} shows how
rotation can be fully disentangled from the mixed mode pattern,
even in the most complex cases where rotational splittings and
period spacings largely overlap \citep[see for instance Figs.
A.4-A.6 of][]{2012A&A...548A..10M}. We then investigate the
principle of the full automation of the analysis of the red giant
oscillation spectrum and test the influence of structure glitches
in Section \ref{glitch}.

A companion paper, \cite{vrard}, presents the automated method for
deriving the asymptotic period spacing in \Kepler\ red giants,
which is based on the present work.

\section{Mixed mode period spacing\label{egalite}}

Following the work of \cite{1979PASJ...31...87S} and
\cite{1989nos..book.....U}, asymptotic expansions of mixed modes
have been derived for different seismic parameters:
eigenfrequencies \citep{2012A&A...540A.143M}, period spacings
\citep{2012ASPC..462..503C}, rotational splittings
\citep{2013A&A...549A..75G,2015A&A...580A..96D}.

Here, we intend to provide expansions taking into account complex
expressions of the pure pressure and pure gravity contributions.

\subsection{Asymptotic expansion}

\cite{1979PASJ...31...87S} and \cite{1989nos..book.....U} derived
an implicit asymptotic relation for mixed modes, which expresses
as
\begin{equation}\label{eqt-asymp}
    \tan\thetap = q \tan\thetag ,
\end{equation}
where $q$ is the coupling factor between the gravity and pressure
components of the modes and where the phase $\thetap$ and
$\thetag$ refer, respectively, to the pressure- and gravity-wave
contributions. At first order, the phases  for dipole modes are
related to the period spacing $\Tg$ and large separation $\Dnu$,
respectively. They write
\begin{eqnarray}
% \nonumber to remove numbering (before each equation)
  \thetag &=& \pi \left( {1 \over \nu \Tg} - \epsg\right) \label{eqt-g0}, \\
  \thetap &=& \pi \left( {\nu\over \Dnu} - {1\over 2} - \epsp\right) \label{eqt-p0} ,
\end{eqnarray}
as used for red giants by \cite{2012A&A...540A.143M}, but here
correctly accounting for the contribution $\ell/ 2$, with
$\ell=1$, and for the offsets $\epsg$ and $\epsp$.

Owing to the form of the mixed-mode asymptotic expansion, we can
modify the phases in Eqs.~(\ref{eqt-g0}) and (\ref{eqt-p0}) with
additional terms, provided their contributions are multiples of
$\pi$. This is a useful artifice for introducing the
eigenfrequencies of either pure pressure or pure gravity modes.
Therefore, we rewrite the phases as
\begin{eqnarray}
% \nonumber to remove numbering (before each equation)
  \thetag &=& \pi {1 \over \Tg}  \left({\displaystyle{1\over\nu}
  -\displaystyle{1\over\nug}}\right), \label{eqt-g} \\
  \thetap &=& \pi {\nu-\nup\over \Dnup} \label{eqt-p},
\end{eqnarray}
where $\nup$ and  $\nug$ are the asymptotic frequencies of pure
pressure and gravity modes, respectively, and $\Dnup$ is the
frequency difference between the consecutive pure pressure radial
modes with radial orders $\np$ and $\np+1$. The offsets and
$\ell/2$ term introduced in Eqs.~(\ref{eqt-g0},\ref{eqt-p0}) are
included in the terms $\nup$ and $\nug$. This means that we now
have the possibility to use Eqs.~(\ref{eqt-g},\ref{eqt-p}) at any
order of the asymptotic expansions for the pure pressure and
gravity contributions.

For $\nug$, acceptable fits of red giant oscillation spectra are
based on the first-order asymptotic expansion
\citep{1980ApJS...43..469T}. For dipole modes, we have
\begin{equation}\label{eqt-g-1erordre}
    {1 \over \nug} = (- \ng+\epsg)\, \Tg ,
\end{equation}
where $\ng$ is the gravity radial order, usually defined with a
negative value, and the offset $\epsg = 1/4 - {\epsg}'$. The
gravity offset ${\epsg}'$ is a small but complicate function
sensitive to the stratification near the boundary between the
radiative core and the convective envelope
\citep{1986A&A...165..218P}. If desirable, we may consider in
Eq.~(\ref{eqt-g}) a second-order expansion for the gravity
expansion \citep{1980ApJS...43..469T}, or the contribution of
buoyancy glitches
\citep[e.g.,][]{2008MNRAS.386.1487M,2015ApJ...805..127C}. This is
done in Section \ref{glitch}.

For $\nup$, the high quality of the seismic data requires the use
of the second-order asymptotic expansion
\citep{2013A&A...550A.126M}. For dipole modes, we have
\begin{equation}\label{eqt-def-nu2eordre}
    \nu\ind{p} (\np) =   \left( \np + {1\over 2} + \epsp + {\alpha
    \over 2}
    (\np-\nmax)^2
    + d_{01}
    \right)  \ \Dnu ,
\end{equation}
where $\np$ is the pressure radial order, $\alpha$ represents the
curvature of the radial oscillation pattern, $\nmax = \numax /
\Dnu - \epsp$ is the non-integer order at the frequency $\numax$
of maximum oscillation signal, and $d_{01}$ is the small
separation, namely the the distance, in units of $\Dnu$, of the
pure pressure dipole mode compared to the midpoint between the
surrounding radial modes \citep{2013A&A...550A.126M}. In this
case, the observed large separation $\Dnup$ in Eq.~(\ref{eqt-p})
increases with increasing radial order:
\begin{equation}\label{eqt-def-dnu2eordre}
    \Dnup =  \Dnu  \ \bigl( 1 + \alpha
    (\np-\nmax) \bigr),
\end{equation}
which is obtained from the derivation of
Eq.~(\ref{eqt-def-nu2eordre}). As for the gravity contribution,
the asymptotic expressions of the pure p modes can be as precise
as required; $\nup$ may include the glitch component due to the
second ionization of helium
\citep{2010A&A...520L...6M,2014MNRAS.440.1828B,2015A&A...579A..84V}.

\begin{figure}
\includegraphics[width=8.8cm]{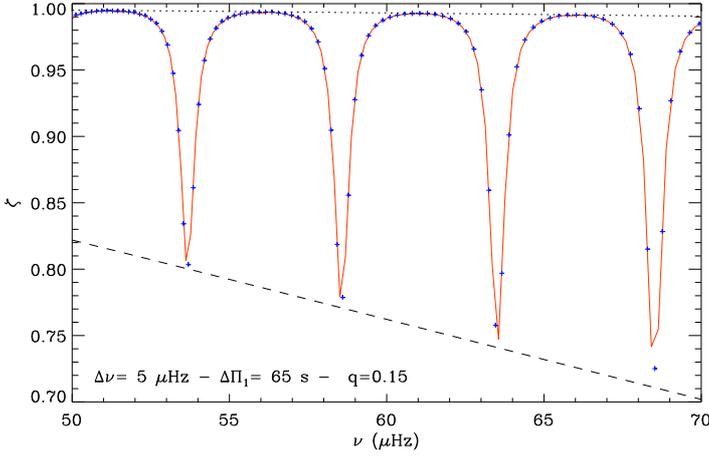}
\caption{Relative period spacings $\DP / \Tg$ and relative
rotational splittings $\zeta$ for a synthetic RGB mixed-mode
spectrum. The red line shows the asymptotic $\zeta$ function; the
blue crosses represent the period spacings derived from the
asymptotic expansion of mixed modes and scaled to $\Tg$. The
dashed line shows the minimum values $\zeta\ind{min}$ reached by
pressure-dominated mixed modes; the dotted line shows the maximum
values $\zeta\ind{max}$ reached by gravity-dominated mixed
modes.}\label{fig-compar}
\end{figure}

\subsection{Bumped period spacing}

In order to retrieve the relative period spacing $\DP/\Tg$ between
two consecutive mixed modes, one needs to introduce the mixed-mode
radial order, $\nm=\np+\ng$. We then choose to write the period
$P=1/\nu$ of a mixed mode as
\begin{equation}\label{eqt-periode}
    P = \nm \Tg + p,
\end{equation}
where the residual term $p$ expresses the departure to an
evenly-spaced comb function. This period is introduced in the
phases $\thetag$ and $\thetap$ provided by Eqs.~(\ref{eqt-g}) and
(\ref{eqt-p}) and linked by the asymptotic relation
(Eq.~\ref{eqt-asymp}). In the gravity contribution, all terms
multiple of $\pi$ cancel, owing to the property of the tangent
function, so that with Eq.~(\ref{eqt-periode}) the term
$\tan\thetag$ reduces to $\tan\pi( p/\Tg - \epsg)$. The derivation
of Eq.~(\ref{eqt-asymp}) with respect to $\nm$ then gives
\begin{equation}\label{eqt-deriv}
    {\displaystyle{\diff\over\diff\nm} \left( \displaystyle{1\over \nm\Tg + p}
    \right)
    \over
    \Dnup \cos^2 \thetap}
    =
    q\
    {\displaystyle{\diff p\over\diff\nm}
    \over
    \Tg \cos^2 \thetag}
    ,
\end{equation}
when the variation of $\Dnup$ with frequency is neglected in the
pressure term.
\begin{figure}
\includegraphics[width=8.8cm]{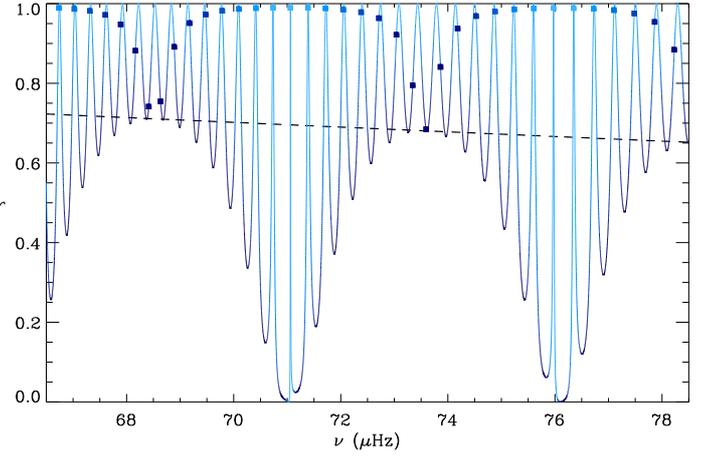}
\caption{Comparison of the function $\zeta$ considered as a
continuous function of frequency (continuous line with a color
modulated by the value of $\cos^2\thetag$) with the values derived
for the mixed-mode frequencies (full squares with a color
modulated by the value of $\cos^2\thetap$). Values of $\cos^2$
close to 0 (1) appear in light blue (dark blue). For
gravity-dominated mixed modes, the phase term $\thetap$ acts for
having $\zeta$ values close to the local maxima; for
pressure-dominated mixed modes, the $\zeta$ values are close to
the local minima $\zeta\ind{min}$ (dashed line). The simulation is
based on the second-order asymptotic pressure-mode pattern
proposed for red giants by \cite{2013A&A...550A.126M}, with $\Dnu=
5\,\mu$Hz, and on the first-order asymptotic expansion of gravity
modes, with $\Tg =65$\,s and $\epsg=0$.}\label{fig-zeta}
\end{figure}
One easily obtains
\begin{equation}\label{eqt-deriv2}
    - \nu^2 {\diff P\over\diff\nm}
    =
    q\ {\Dnup \over \Tg }\ {\cos^2\thetap \over \cos^2\thetag}\
    \left[ {\diff P\over\diff\nm} - \Tg\right]
    ,
\end{equation}
hence
\begin{equation}\label{eqt-deriv3}
    {1\over \Tg}  {\diff P\over\diff\nm}
    =
    \left[1 +
    {1 \over q} { \nu^2 \Tg \over \Dnup }
    {\cos^2\thetag \over \cos^2\thetap}
    \right]^{-1}.
\end{equation}
The case $\diff\nm=1$ corresponds to the period difference between
two consecutive mixed modes, and thus provides us with the
relative bumped period spacing
\begin{equation}\label{eqt-zeta-P}
    {\DP \over  \Tg} = \zeta(\nu),
\end{equation}
where the definition
\begin{equation}\label{eqt-zeta}
    \zeta(\nu) = \left[1+ {1\over q}
     {\nu^2 \Tg \over \Dnup}
    {\cos^2 \pi \displaystyle{1\over \Tg}
    \left(\displaystyle{{1\over \nu} - {1\over\nug}}\right)
    \over
    \cos^2 \pi \displaystyle{\nu-\nup\over \Dnup}}
    \right]^{-1} .
\end{equation}
matches the first-order asymptotic expansion derived by
\cite{2015A&A...580A..96D} for expressing the mixed-mode
rotational splitting
\begin{equation}\label{eqt-zeta-rotations}
     \dnurot =
     {\dnurot}\ind{,g}\ \zeta
     +
     {\dnurot}\ind{,p}\ (1-\zeta)
     ,
\end{equation}
where ${\dnurot}\ind{,g}$ and ${\dnurot}\ind{,p}$ are the
rotational splittings related to pure gravity or pure pressure
modes. In the limit case where the mean envelope rotation is
negligible compared to the mean core rotation, this simplifies
into
\begin{equation}\label{eqt-zeta-rotation}
     \zeta  (\nu) \simeq
     {\dnurot \over {\dnurot}\ind{,g} }
     .
\end{equation}
Hence, we have demonstrated that the period spacing compared to
$\Tg$ (Eq.~\ref{eqt-zeta-P}) and the rotational splitting compared
to ${\dnurot}\ind{,g}$ (Eq.~\ref{eqt-zeta-rotation}) follow the
same distribution. This result emphasizes that the coupling
between the pressure and gravity terms plays the same role for
arranging the mixed-mode periods and the (frequency) rotational
splittings. As shown in \cite{2013A&A...549A..75G} and
\cite{2015A&A...580A..96D}, this behavior is governed by the
inertia of the mixed modes: $\zeta$ is the ratio between the
kinetic energy in the radiative cavity and the total kinetic
energy. The equality between $\DP / \Tg$ and $\zeta$ is seen in
Fig.~\ref{fig-compar}, where the bumped period spacings $\DP$
defined as $P(\nm+1)-P(\nm)$ are compared to the function $\zeta$.
In order to avoid quantization biases, we have plotted $\DP / \Tg$
at the abscissae $(P(\nm)+P(\nm +1))/2$.

We must stress that the values of $\zeta$ obtained for the
discrete mixed-mode frequencies $\nu$ do not reflect the
complexity of the function $\zeta$ if considered as a continuous
function of frequency. For clarity, we denote $\zeta(\nu)$ the
discrete values and $\zeta(f)$ the continuous function. The
contributions of the phases $\thetap$ and $\thetag$ to $\zeta(f)$
are shown in Fig.~\ref{fig-zeta}.

Interestingly, the global pattern $\zeta(\nu)$ reached for mixed
modes mainly depends on the asymptotic parameters of the pure
pressure modes, since the location of the local minima only
depends on the frequency of the pure pressure dipole modes.
Conversely, it hardly depends on $\Tg$ and $q$ \citep{vrard}.
Limiting cases corresponding to gravity-dominated mixed modes (g-m
mode) and pressure-dominated mixed modes (p-m mode) are derived in
the Appendix.

\subsection{Comparison with previous work}

Recently, \cite{2014ApJ...781L..29B} have remarked that the ratio
$I_1/I_0$ of the inertia of dipole modes compared to radial modes
may provide an alternative to using mixed-mode frequencies and
give essentially the same information as the mixed-mode asymptotic
equation. We confirm that this is the case: their function
$I_1/I_0$ can be related to $\zeta$ if one assumes that the
difference between the total inertia of mixed modes and the
contribution of the radiative cavity only exactly corresponds to
the inertia of radial modes. Their equation (10), with  the factor
$q^2/4$ corrected into $q$, can be written
\begin{equation}\label{eqt-zeta-inerties}
    {I_1\over I_0} = {1\over 1 - \zeta}
    .
\end{equation}
A different expression of the bumped period spacing was earlier
introduced by \cite{2012ASPC..462..503C}. Rewriting Eq. (34) of
his work with our notation gives
\begin{equation}\label{eqt-jcd-34}
    {\DP \over  \Tg} =  \left[1 -
    {\nu^2 \Tg \over \pi}
    {\diff \Phi\over
    \diff\nu}
    \right]^{-1},
\end{equation}
with $\Phi$ defined in \cite{2012ASPC..462..503C} by
\begin{equation}\label{eqt-jcd-phi}
    \tan\Phi = {q\over \tan \thetap}.
\end{equation}
By comparison with Eq.~(\ref{eqt-asymp}), we get $\Phi \equiv
\pi/2- \thetag$. The term $\diff \Phi / \diff\nu$ derived from
Eq.~(\ref{eqt-jcd-phi}) ensures the agreement between the
different expressions of $\DP / \Tg$. We note that the use of
Eq.~(\ref{eqt-zeta}) avoids the introduction of the derivative
term present in Eq.~(\ref{eqt-jcd-34}). As rapid variations arise
from the $\thetag$ contribution, our new expression of $\DP/ \Tg$
(Eqs.~\ref{eqt-zeta-P}, \ref{eqt-zeta}) provides a more precise
numerical result than Eq.~(\ref{eqt-jcd-34}) requiring numerical
derivative.

%............................................................................
\section{Stretched periods\label{applications}}

In this section, we show how the function $\zeta$ can be derived
from the analysis of the pressure mode pattern and how it can be
used to extract information on the pure gravity mode pattern.

\subsection{Identification of the mixed modes}

In a red giant oscillation spectrum, the location of the mixed
modes is in fact fixed by the measurement of the large separation
$\Dnu$. The determination of $\Dnu$, first derived from the
envelope autocorrelation function \citep{2009A&A...508..877M},
then refined with the universal red giant oscillation pattern
\citep{2011A&A...525L...9M}, provides the identification of the
radial modes and helps locating the frequency ranges where mixed
modes cannot be mistaken for radial or quadrupole modes. The
background parameters, derived as in \cite{2012A&A...537A..30M},
are used for correcting the granulation contribution in the
frequency range around $\numax$ where oscillations are observed.
Hence, mixed modes can be automatically identified in frequency
ranges having no radial and quadrupole modes, with a height
significantly above the background.

\begin{figure}
\includegraphics[width=8.8cm]{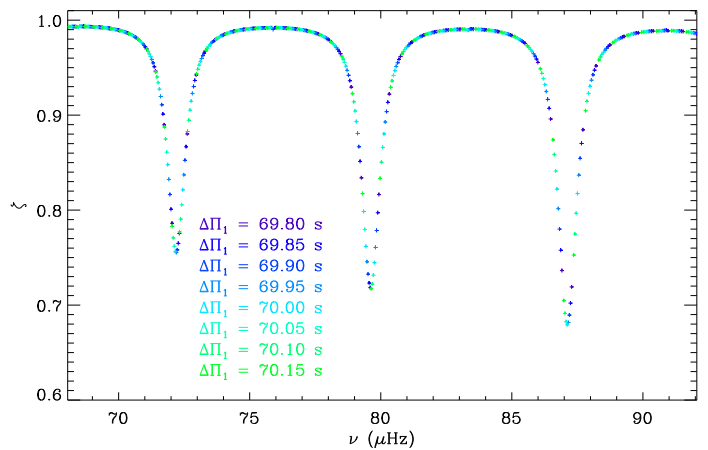}
\caption{Interpolation of the function $\zeta$ for the period
spacing $\Tg$, derived from the values obtained with various
periods  in the range $\Tg (1\pm \numax \Tg /2)$, where $\numax$
is the frequency of maximum oscillation
signal.}\label{fig-zeta-interp}
\end{figure}

\begin{figure}
\includegraphics[width=8.8cm]{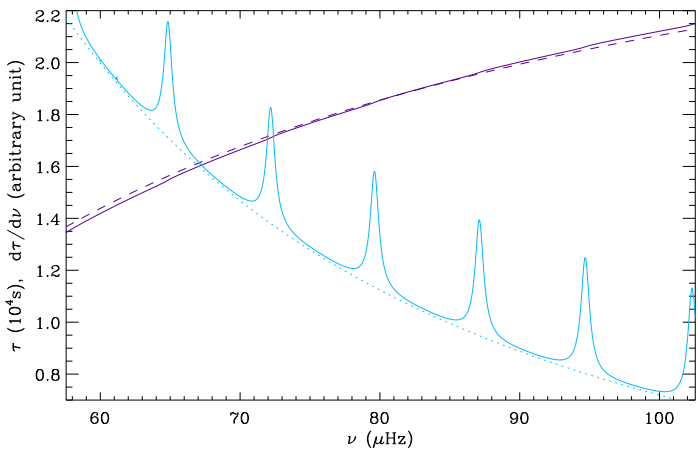}
\caption{Stretched period $\tau$ as a function of frequency (dark
blue line), compared to a reference varying as $1 /\nu$ (dashed
line). The derivative $\diff\tau / \diff\nu$ is superimposed
(light blue line), with an arbitrary scaling factor to fit in the
same window as $\tau$; the dotted line is representative of the
term $1/\nu^2$.}\label{fig-tau}
\end{figure}

\subsection{Estimate of the function $\zeta$}

The same information is used for deriving the function $\zeta$.
The precise location of the minima of the function is obtained
from the location of the p-m modes, which is fixed by the radial
pattern. Then, basic information from the $\Tg$ - $\Dnu$ diagram
\citep{2014A&A...572L...5M} is enough to obtain a precise estimate
of $\zeta$. Large modifications of $\Tg$ only slightly modify the
$\zeta$ profile \citep{vrard}. Furthermore, the variation of $q$
with $\Tg$, hence with $\Dnu$, also obeys a deterministic
relation, so that $q$ can be predicted with a precision better
than 20\,\%, which is enough.

As shown in Fig.~\ref{fig-zeta}, the function $\zeta$ is properly
defined for the mixed-mode frequencies only. Therefore, to obtain
a continuous function representative of the period change due to
the mixing of the modes, we have to interpolate the values of
$\zeta(\nu)$. Figure \ref{fig-zeta-interp} graphically explains
how this can be obtained with a small modulation of the period
spacing, assuming that $\zeta$ only slightly depends on $\Tg$.
This property is fully developed in \cite{vrard}.

\subsection{Stretched periods}

The interpolated form of $\zeta$ is then used to turn the
frequencies into periods $\tau$ with the differential equation
\begin{equation}\label{eqt-stretch}
    \diff\Pm = {1\over \zeta} {\diff \nu \over \nu^2}
    ,
\end{equation}
from which we can integrate the periods $\tau$ of mixed modes
(Fig.~\ref{fig-tau}). The constant of integration can be
arbitrarily fixed since it plays no role. Owing to the $\zeta$
profile (Fig.~\ref{fig-compar}), we call them \emph{stretched
periods}. They are used for drawing \'echelle diagrams, where
$\nu$ is plotted as a function of $\tau$ modulo $\Tg$.

\subsection{\'Echelle diagrams based on the stretched periods}

\begin{figure*}
\includegraphics[width=7.6cm]{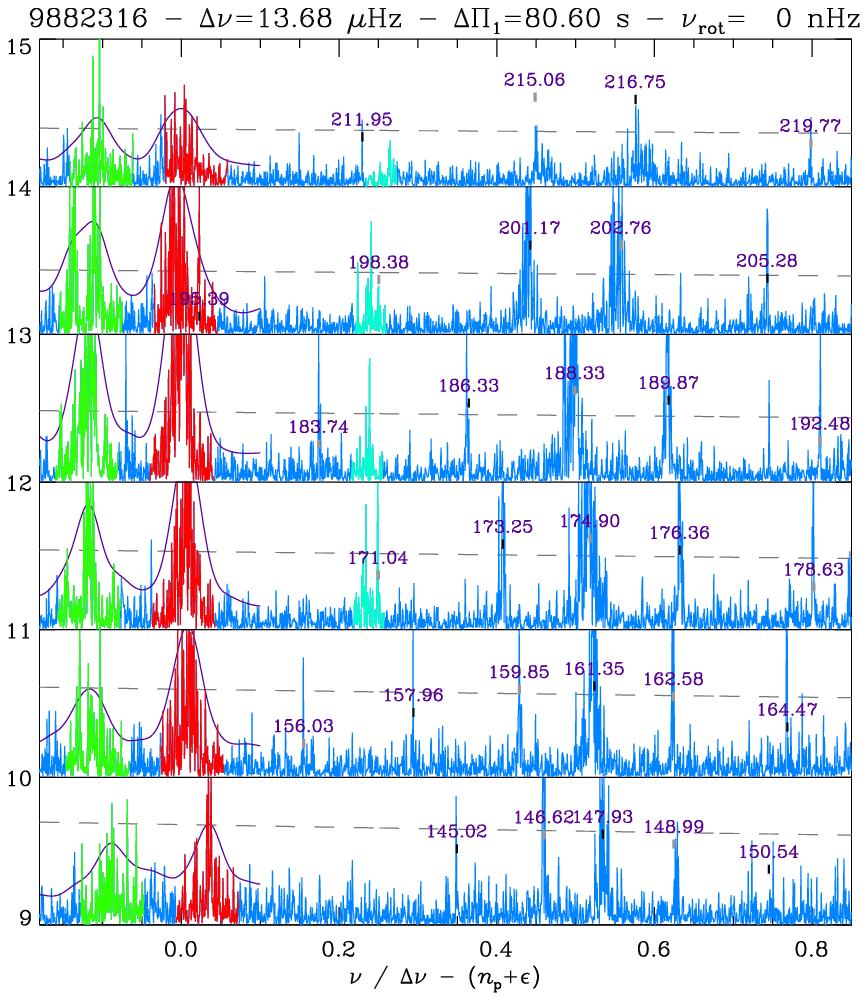}
\includegraphics[width=5.0cm]{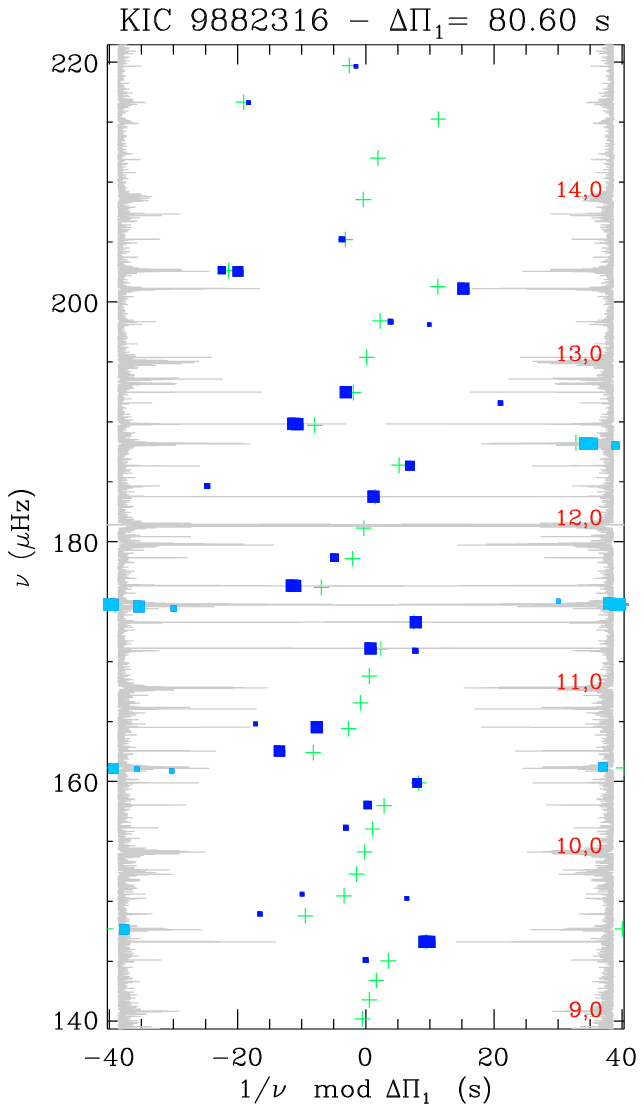}
\includegraphics[width=5.0cm]{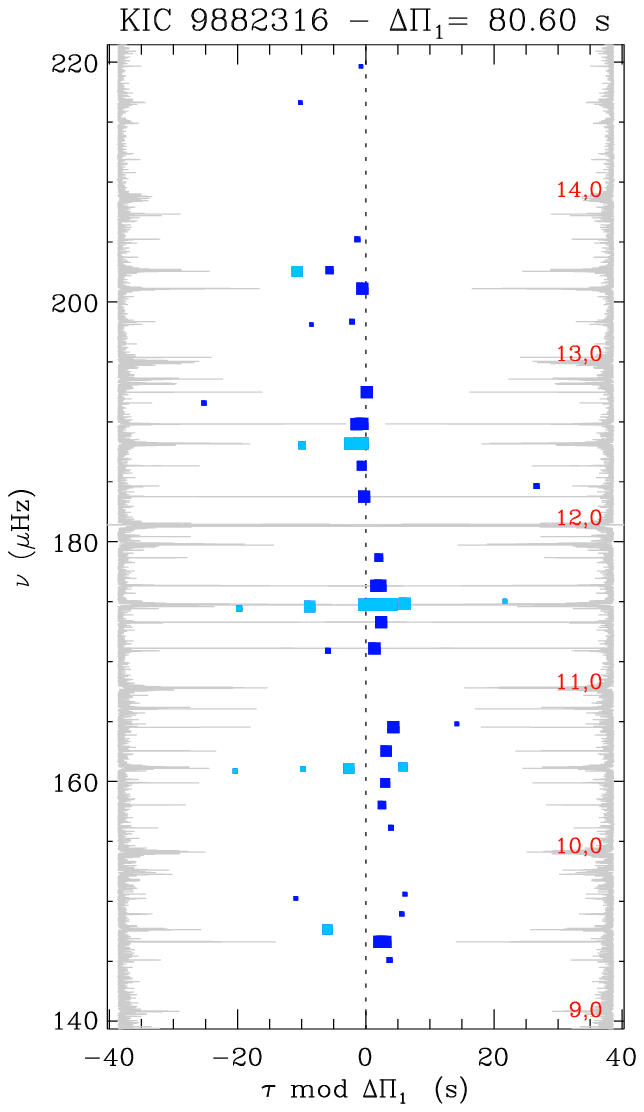}
\caption{\'Echelle diagrams of the RGB star KIC 9882316 that shows
no rotational signature.
 {\textsl{Left}:} Classical frequency \'echelle diagram  as a
function of the dimensionless frequency $\nu/\Dnu -
(\np+\varepsilon)$. The radial order $\np$ is indicated on the
y-axis. Radial modes (highlighted in red) are centered on 0,
quadrupole modes (highlighted in green), near $-0.12$ (with a
radial order $\np-1$), and $\ell=3$ modes, sometimes observed,
(highlighted in light blue) near 0.22. Dipole mixed modes are
identified with the frequency given by the asymptotic relation of
mixed modes, in $\mu$Hz. The fit is based on  peaks showing a
height larger than eight times the mean background value (grey
dashed lines).
 {\textsl{Middle}:} Classical period \'echelle diagram, where the
abscissa is the period modulo the period spacing $\Tg$. The most
prominent mixed modes, marked with blue filled squares (in light
blue for peaks in the vicinity of p-m modes), are automatically
identified. In the background of the figure, the spectra are
plotted twice and top to tail for making the mode identification
easier, with the pressure radial orders indicated on the radial
modes.
{\textsl{Right}:} Stretched period \'echelle diagram, where the
abscissa is the stretched period $\tau$ modulo the period spacing
$\Tg$. Pressure-dominated mixed modes are coded in light blue.
}\label{fig-echelle}
\end{figure*}

We first consider the case of a star seen pole-on, for which
rotation is not an issue; the full case including rotation is
solved in Section \ref{disentangling}. The left panel of
Fig.~\ref{fig-echelle} shows the varying period spacings of the
mixed modes. The \'echelle diagram based on the periods $1/\nu$
(middle panel) shows the classical S-shape
\citep{2011Natur.471..608B,2012A&A...540A.143M}, whereas the
\'echelle spectrum based on the stretched periods $\tau$ (right
panel) exhibits a single nearly-vertical ridge, where discrepant
peaks can be easily identified as speckle structures due to the
short-lived p-m modes or as $\ell=3$ modes. As for classical
\'echelle diagrams used for estimating the large separation
$\Dnu$, the value of the period used for folding the stretched
period influences the slope of the ridge. It is easy to ensure a
vertical alignment and then extract the measurement of the period
spacing $\Tg$. This is done in the companion paper \citep{vrard}
and we let the discussion on the performance for measuring $\Tg$
to this work.

\begin{figure*}
\includegraphics[width=5.9cm]{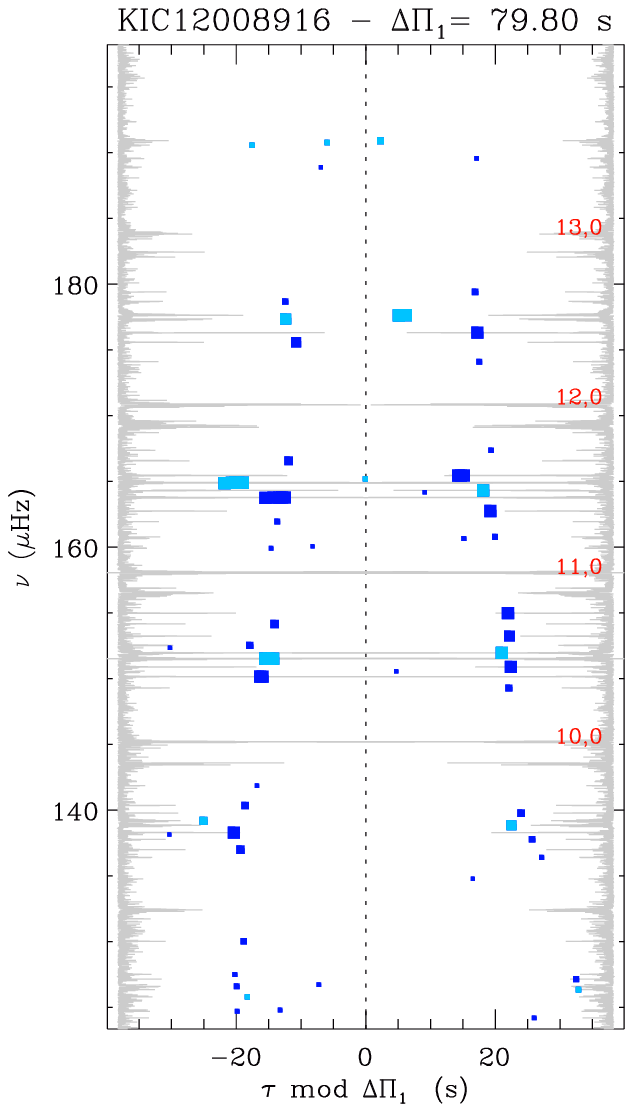}
\includegraphics[width=5.9cm]{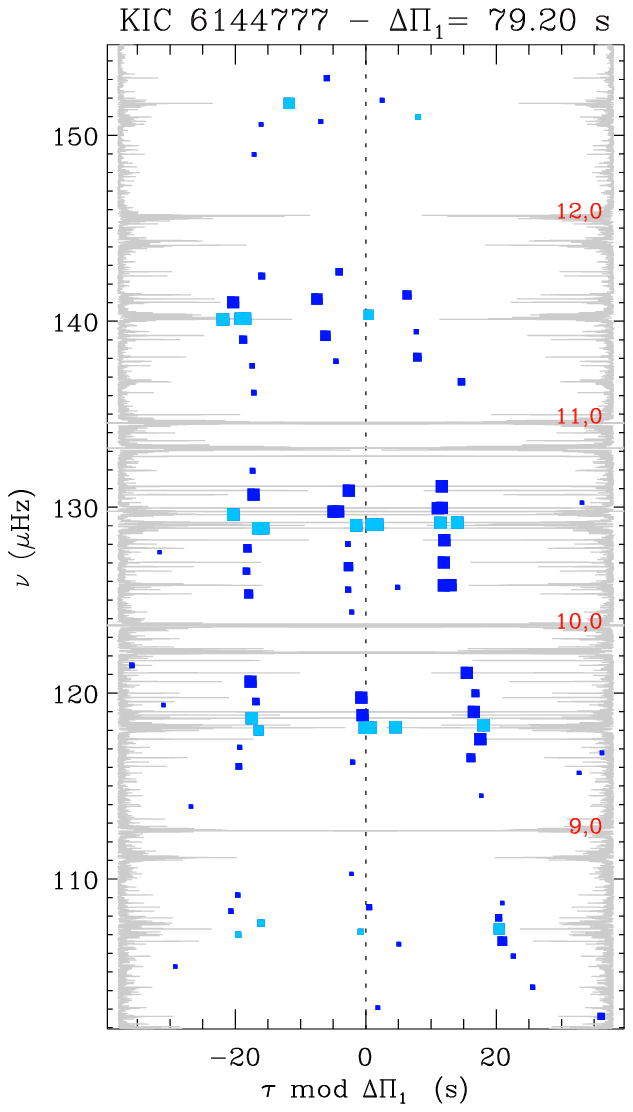}
\includegraphics[width=5.9cm]{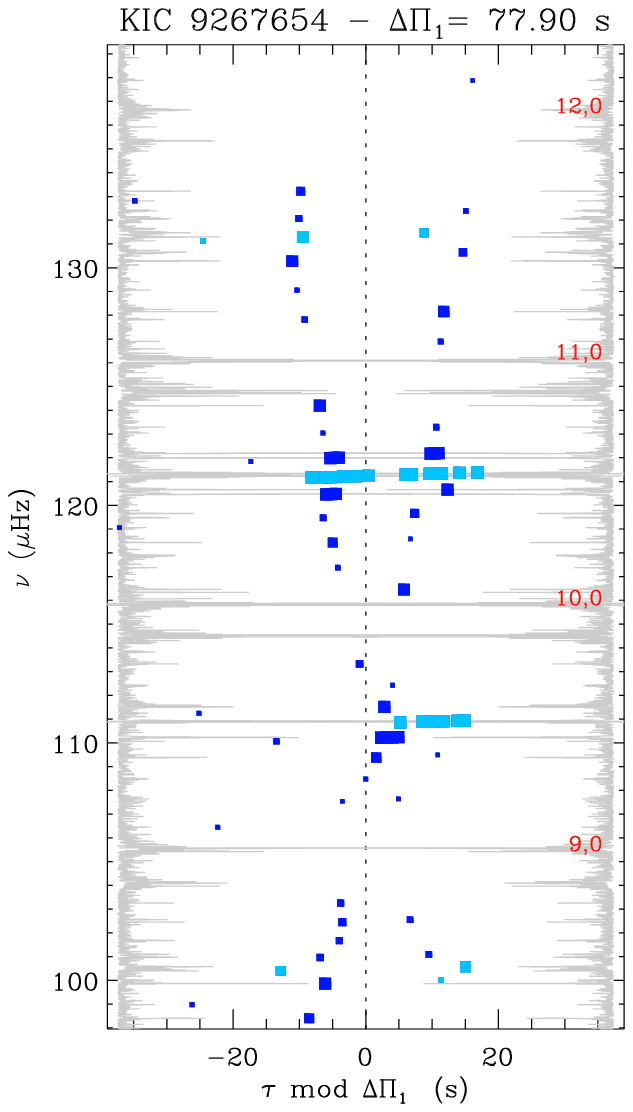}
\caption{\'Echelle diagrams based on the stretched periods of
dipole mixed modes. The abscissae are the periods $\tau$, derived
from Eq.~(\ref{eqt-stretch}), modulo the asymptotic period spacing
$\Tg$; the ordinates are frequencies. The most prominent mixed
modes, marked with filled squares, are automatically identified.
In the background of the figure, the spectra are plotted twice and
top to tail for making the mode identification easier, with the
pressure radial orders indicated on the radial modes. From the
left to the right, we show RGB stars with different rotational
splittings.
{\textsl{Left}:} Star seen equator-on, where only $m=\pm 1$ modes
are visible, and with a moderate mean core rotation; they draw two
ridges from which the rotation frequency $\dnurot$ can be derived.
{\textsl{Middle}:} Star seen with an inclination such that all
three components of the rotational multiplets are visible, and
with a moderate mean core rotation.
{\textsl{Right}:} Star seen equator-on; as a consequence of the
large mean core rotation, the ridges $m=\pm 1$ intersect.
}\label{fig-rotation}
\end{figure*}

\subsection{Precise measurement of the coupling factor}

The precision on the measurement of the coupling term $q$ can be
significantly enhanced. Previous estimates showed uncertainties as
large as 0.05 for a mean value of $q$ about 0.2
\citep{2012A&A...540A.143M}. Here, the expression of
$\zeta\ind{min}$ gives a new way for measuring $q$ in an automated
manner. As in Figure~\ref{fig-compar}, one can compare the period
spacings, for a fixed $m$ ridge, with the function $\zeta\ind{min}
\Tg$, and derive $q$ with a smaller uncertainty. It is even
possible to reduce this uncertainty to about 0.02 by tuning $q$
for ensuring a minimum spread around p-m modes in the \'echelle
diagram constructed with the stretched periods.

\section{Disentangling rotation\label{disentangling}}

In this section, we intend to use the analytical expression of the
period spacing (Eq.~\ref{eqt-zeta}) to disentangle the complex
structure of the mixed mode pattern, especially for red giants on
the RGB where the most complex cases are met
\citep{2012A&A...548A..10M}. For this purpose, we have considered
stars analyzed in \cite{2012Natur.481...55B} and
\cite{2012A&A...537A..30M,2012A&A...540A.143M,2012A&A...548A..10M}.
We have retrieved the public data set in the \Kepler\ archive in
order to benefit from the longest available time series (about 4
years).

\begin{table*}[t]
\caption{Rotational splittings\label{table-rotation}}
\begin{tabular}{rcccccccccc}
 \hline
      &\multicolumn{2}{c}{Pressure modes}&\multicolumn{3}{c}{Hand-made analysis}
      &\multicolumn{5}{c}{Analysis based on stretched periods}\\
 KIC  &$\numax$&$\Dnu$&$\Tg$&$\nmix (\numax)$& $\dnurotmax^{(a)}$&$\DPmp$&$\DPmm$
      &$\langle \DPm \rangle$&  $\xrot$ & $\dnurotmax^{(b)}$ \\
      &$\mu$Hz &$\mu$Hz&s &   & nHz & s & s & s & \% & nHz\\
 \hline
%  3744043&     &      &     &   & ? \\
%  5858947&     & 14.47&84.3 &   & 515\\
  6144777&128.4&11.029&79.0 &8.5& 242 & 78.75 &  79.25 & 79.0 & 0.34 & 227 \\
%  6928997&     &      &     &   & ? \\
  9267654&118.0&10.358&78.0 &9.5& 490 & 77.30 &  78.55 & 77.9 & 0.75 & 523 \\
%  9574650&     & 9.63 &76.0 &   & 365\\
% 10777816&     &      &     &   &  \\
% 10866415&     & 8.75 &75.2 &   &  355 \\
% 11550492&     & 8.70 &74.9 &   &  390 \\
 12008916&160.8&12.906&79.8 &6.3& 430 & 79.45 &  80.15 & 79.8 & 0.46 & 409 \\
\hline
\end{tabular}

(a) rotational splitting derived from the use of
\cite{2013A&A...549A..75G}.

(b) rotational splitting derived from this work
\end{table*}

\subsection{Rotational splittings\label{full-id}}

The rotation being considered as a perturbation of the dipole
mixed-mode pattern and the surface rotation being negligible in
most cases, the unperturbed frequency $\nu$ is changed by rotation
into $\nu - m \dnurot$ \citep{2013A&A...549A..75G}. According to
Eq.~(\ref{eqt-zeta-rotations}) and to the profile of $\zeta$, this
reduces to $\nu - m \zeta \dnurotmax$, even in the cases where the
envelope rotation cannot be neglected, as for instance in the
secondary clump \citep{2015A&A...580A..96D}. For dipole modes,
azimuthal orders $m \in \{-1,0,1\}$. Hence, the differences in the
stretched periods of dipole modes with same azimuthal order $m$
can be approximated by
\begin{equation}\label{eqt-spacing-rot}
    \DPm = \Tg \ \left(1+ 2\, m \,{\zeta \, \dnurotmax  \over
    \nu}\right) ,
\end{equation}
if we assume that the mean envelope rotation can be omitted
\citep{2013A&A...549A..75G}. Even for an important rotational
splitting, as large as or even larger than the period spacing, the
relative correction due to rotation is small. Therefore, the
stretched period spacings ${\Delta\Pm}_{m = \pm 1}$ are close to
$\Tg$ (Table \ref{table-rotation}). Since modes are observed in a
limited frequency range around the frequency $\numax$ of maximum
oscillation signal, we can approximate the varying term
$\zeta/\nu$ of Eq.~(\ref{eqt-spacing-rot}) by $\nmix
/((\nmix+1)\numax)$, where the function
\begin{equation}\label{eqt-nmix}
    \nmix (\nu) = {\Dnu \over \Tg \nu^2}
\end{equation}
shows how the period spacings compare to the large separation
$\Dnu$. It corresponds to the number of gravity modes per
$\Dnu$-wide frequency range around the frequency $\nu$. This shows
that the period spacing of each component of the rotation
multiplet is nearly uniform, close to
\begin{equation}\label{eqt-spacing-rot2}
    \DPm  \simeq  \Tg \ ( 1 + m\, \xrot ) ,
\end{equation}
with
\begin{equation}\label{eqt-prop}
    \xrot = 2 \ {\nmix (\numax) \over \nmix (\numax) +1} {\dnurotmax \over
    \numax}
    .
\end{equation}
We note that the ratio $ \nmix / (\nmix+1)$ is a consequence of
the addition of an extra pressure mode among the $\nmix$ gravity
modes that construct the $(\nmix +1)$ mixed modes in a $\Dnu$-wide
frequency range.

\subsection{Measurement of $\Tg$ and $\dnurotmax$}

The use of Eq.~(\ref{eqt-spacing-rot}) allows us to correct the
variation of the rotational splitting with frequency and their
asymmetry near the p-m modes. According to
Eq.~(\ref{eqt-spacing-rot2}), each azimuthal order forms a
well-identified ridge in the \'echelle diagram. Each ridge
approximately shows linear variation of the rotationally perturbed
period spacings (Fig.~\ref{fig-rotation}), from which we can
derive the period spacing $\Tg$ and the rotational splitting
$\dnurot$. When the rotation splitting exceeds half the mixed-mode
spacing, the $m=\pm1$ ridges intersect (Fig.~\ref{fig-rotation},
right panel). Results obtained for these stars are given in
Table~\ref{table-rotation}. Values agree with the rotational
splittings measured in \cite{2012A&A...548A..10M}, with a small
difference due to the approximation used in this earlier paper,
which is explained by the difference between the approximate form
of the rotational splitting found by \citep{2013A&A...549A..75G}
and the more precise asymptotic rotational splitting
\citep{2015A&A...580A..96D}.

When the surface rotation cannot be omitted, as is the case on the
low RG \citep{2012ApJ...756...19D,2014A&A...564A..27D} or in the
secondary red clump \citep{2015A&A...580A..96D}, the rotation
correction $\zeta \dnurot$ in Eq.~(\ref{eqt-spacing-rot}) must be
changed into $\zeta (\dnurotcore - \dnurotenv) + \dnurotenv$
\citep{2013A&A...549A..75G}. Conclusions remain the same, but
estimating the separated contribution of the core and of the
envelope then requires a separated study of p-m and g-m modes.
% In principle, this can be done with the scheme we propose.

\subsection{Toward an automated determination of the global seismic parameters\label{automat}}

The method presented above has shown its ability to exhibit in a
clear way the rotational structure of the mixed modes and identify
their azimuthal orders. This comes from the fact that
Eq.~(\ref{eqt-stretch}) allows us to directly analyze the period
pattern, and helps us avoiding the identification of the mixed
modes prior to the measurement of $\Tg$ as in other methods
\citep[e.g.,][]{2015MNRAS.447.1935D}. In fact, in many cases and
especially on the RGB, defining period spacings of dipole mixed
modes requires first the full identification of the dipole modes,
which is a severe drawback for complex spectra: rotational
splittings, $\ell=3$ modes, and short-lived p-m mode conspire
against the identification of the angular degree and azimuthal
order of the modes, then against the measurement of relevant
$\DP$. The use of Eq.~(\ref{eqt-stretch}) alleviates this problem.

All stages used to draw the \'echelle diagrams were based on
automated methods. The steps for obtaining $\Dnu$ were recalled in
Section \ref{applications} and are not modified by rotation. The
asymptotic period spacings can be obtained in an automated way too
\citep{vrard}. The next step is to make profit of the views
developed in this work to automate the measurement of the
rotational splitting too; this is the aim of a forthcoming paper.
The method can also be used to investigate possible rapid rotators
\citep{2013A&A...554A..80O}. Possible complications arising from
structural glitches are presented in the next Section.

\section{Glitches\label{glitch}}

In this Section, we analyze how structural glitches modify the
mixed-mode pattern and might complicate the oscillation spectrum.
Glitches are due to rapid variation either in the sound-speed
profile or in the \BV\ profile.

\subsection{Sound-speed glitches}

\begin{figure}
\includegraphics[width=4.4cm]{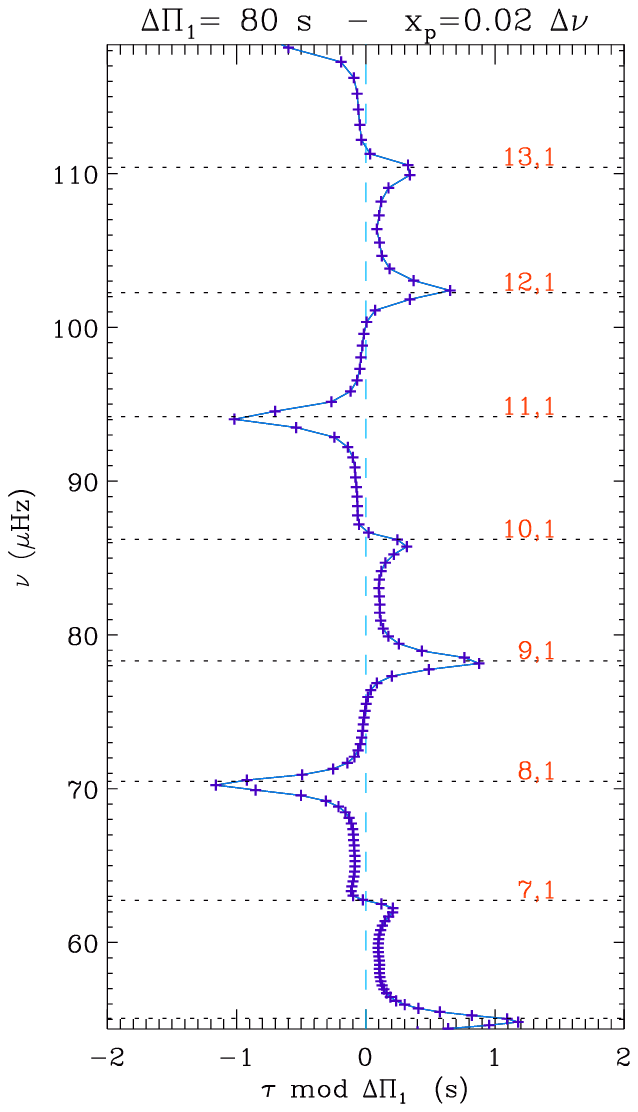}
\includegraphics[width=4.4cm]{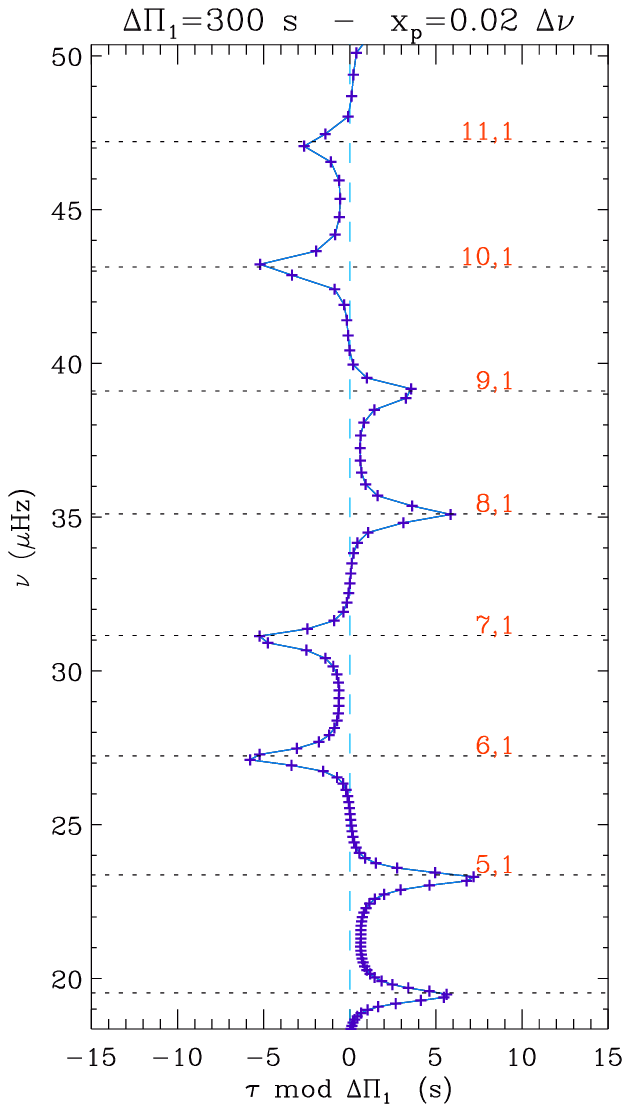}
\caption{Stretched period \'echelle diagrams of synthetic
mixed-mode spectra including sound-speed glitches.
 \textsl{Left:} RBG star;
 \textsl{Right:} red clump star. Horizontal dotted lines indicate the
location of the pure pressure dipole modes.}\label{fig-glitchp}
\end{figure}

The pressure mode spectrum of red giants is modulated by acoustic
glitches due to abrupt variation of the first adiabatic exponent
at the location of the second ionization of helium
\citep{2010A&A...520L...6M,2014MNRAS.440.1828B}. We have modelled
the pressure glitches following \cite{2015A&A...579A..84V}, with a
simple sinusoidal modulation added to the pure pressure modes
$\nup$. We considered two cases, corresponding to typical clump
and RGB stars, and used the mean values of the glitch periods and
amplitudes given in \cite{2015A&A...579A..84V}: the periods of the
glitch signature are in the range [3 -- 4], in units $\Dnu$,  and
the amplitude about a few percents, in units $\Dnu$ too. We note
that the introduction of the sound-speed glitch does not modify
the scheme presented in Section \ref{egalite}, so that one can
easily retrieve the period spacings (Fig.~\ref{fig-glitchp}).

The analysis of the stretched periods confirms that only p-m modes
are significantly affected by the sound-speed glitches. Their
location, hence the value $d_{01}$ of the small separation in
Eq.~(\ref{eqt-def-nu2eordre}), must be precisely determined in
order to perform a correct analysis of the glitch signature.

The influence on the g-m modes is reduced since the pure
gravity-mode pattern is not affected by the sound-speed glitch.
According to the amplitude observed for acoustic glitches, the
deviation in the p-m mode is limited to a small fraction of $\Tg$.
The relative amplitude $x\ind{p}$ of the acoustic glitch, which
corresponds to a frequency shift of $x\ind{p} \Dnu$, translates
into a period shift of the pure pressure modes about $x\ind{p} \,
\nmix \, \Tg$. The maximum period shift $\delta\Pm\ind{p}$ of the
mixed modes is much smaller, since the gravity component of the
wave is not changed. We provide an estimate of this effect with a
set of simulation of synthetic glitches, performed at various
evolutionary stages. Scaling relations of $\delta\Pm\ind{p}$ with
$\Dnu$, $\numax$ and $\Tg$ can be summarized by
\begin{equation}\label{shift-glitchp}
    \delta\Pm\ind{p} \simeq 0.4 \,\, x\ind{p} \, \nmix(\numax)^{1/3} \, \Tg .
\end{equation}
Observed values of $x\ind{p}/\Dnu$ are about 1 - 2\,\%
\citep{2015A&A...579A..84V}, and at $\numax$ there are about 10
gravity modes per $\Dnu$ frequency range, so that the perturbation
$\delta\Pm\ind{p} / \Tg$ is at most about 1 - 2\,\%. In any case,
sound-speed glitches are neither expected to perturb the
determination of the mixed-mode pattern, nor the measurement of
$\Tg$.

\begin{figure}
\includegraphics[width=8.8cm]{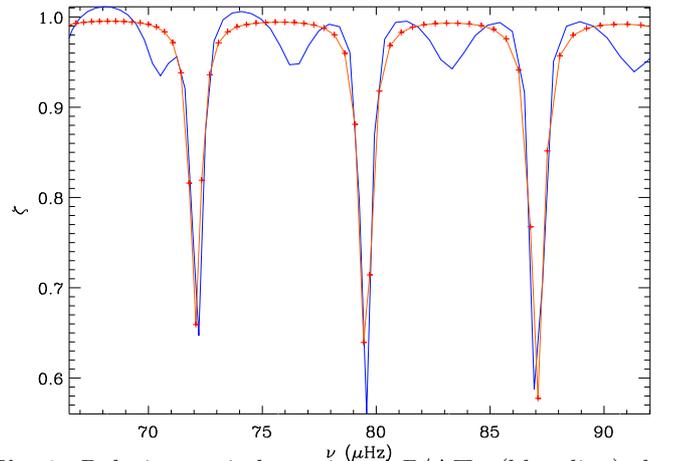}
\caption{Relative period spacing $\DP / \Tg$ (blue line) derived
from the function $\zeta$ calculated without glitches (red line
and plusses). Both values of $\zeta$ and $\zeta - \diff \glitchg
/\diff \nm$ are superimposed and cannot be
distinguished.}\label{fig-compare-zeta-glitch}
\end{figure}

\subsection{Buoyancy glitches}

For buoyancy glitches, we use the analysis of
\cite{2008MNRAS.386.1487M} describing them in SPB and $\gamma$
Doradus stars, since the analytical expressions they derive are
general enough to be applied to the red giant case. The asymptotic
period spacing, modulated by a buoyancy glitch, can be written
\begin{equation}\label{eqt-modul-glitchg}
    {\Tg}\ind{glitch} = \Tg \ \bigl(1+ \glitchg (1/\nu) \bigr)
    ,
\end{equation}
where  $g$ is a periodic function. Its period is defined by
\begin{equation}\label{eqt-periode-glitch}
    \periodeglitch = {\Delta\Pi\ind{g} \over \Tg}
    =
    {
    \displaystyle{\int_{r_{N1}}^{r_{N2}} \NBV  {\diff r \over r}}
    \over
    \displaystyle{\int_{r\ind{g}}^{r_{N2}} \NBV  {\diff r \over r}}
    }
    ,
\end{equation}
where $\NBV$ is the \BV\ frequency, $r_{N1}$ ($r_{N2}$) is the
inner (outer) boundary of the radiative core, and where
${r\ind{g}}$ is the radius of the buoyancy glitch\footnote{We note
that this definition implies that the ratio $\periodeglitch$ is
necessarily larger than unity. It is also possible to change the
boundaries of the denominator into, respectively, $r_{N1}$ and
$r\ind{g}$, for another definition $\periodglitch$ of the period,
which verifies $1/\periodglitch = 1 -  1/ \periodeglitch$.}.

\begin{figure*}
\includegraphics[width=3.5cm]{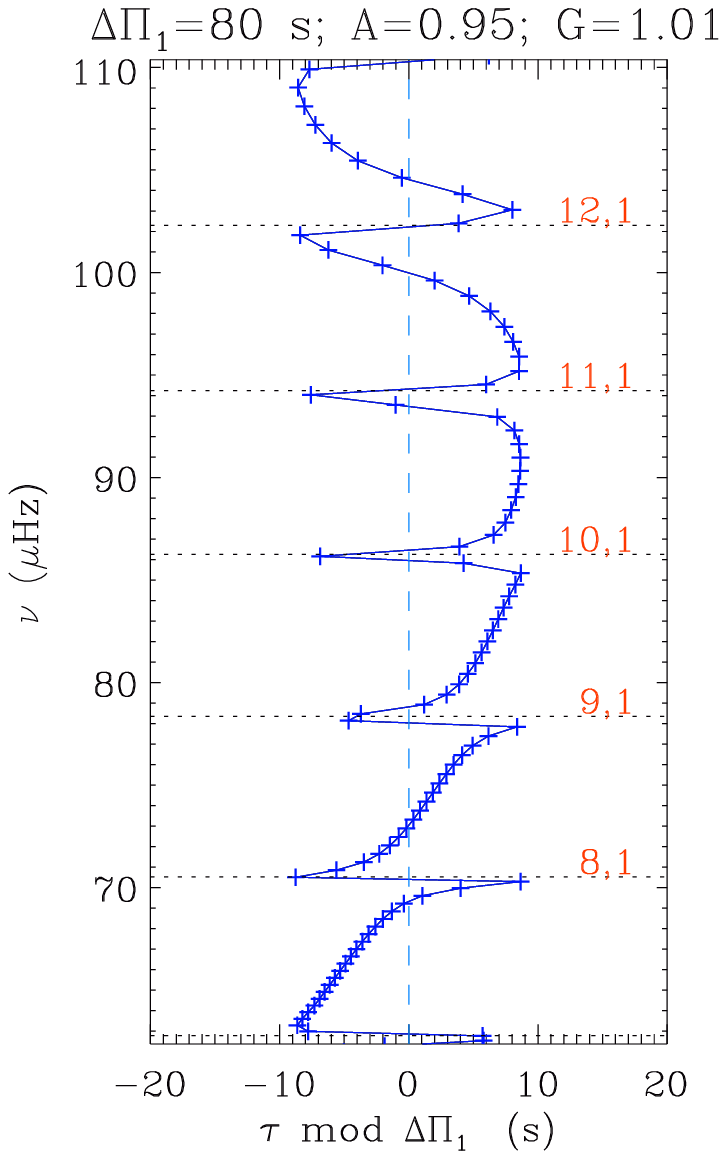}
\includegraphics[width=3.5cm]{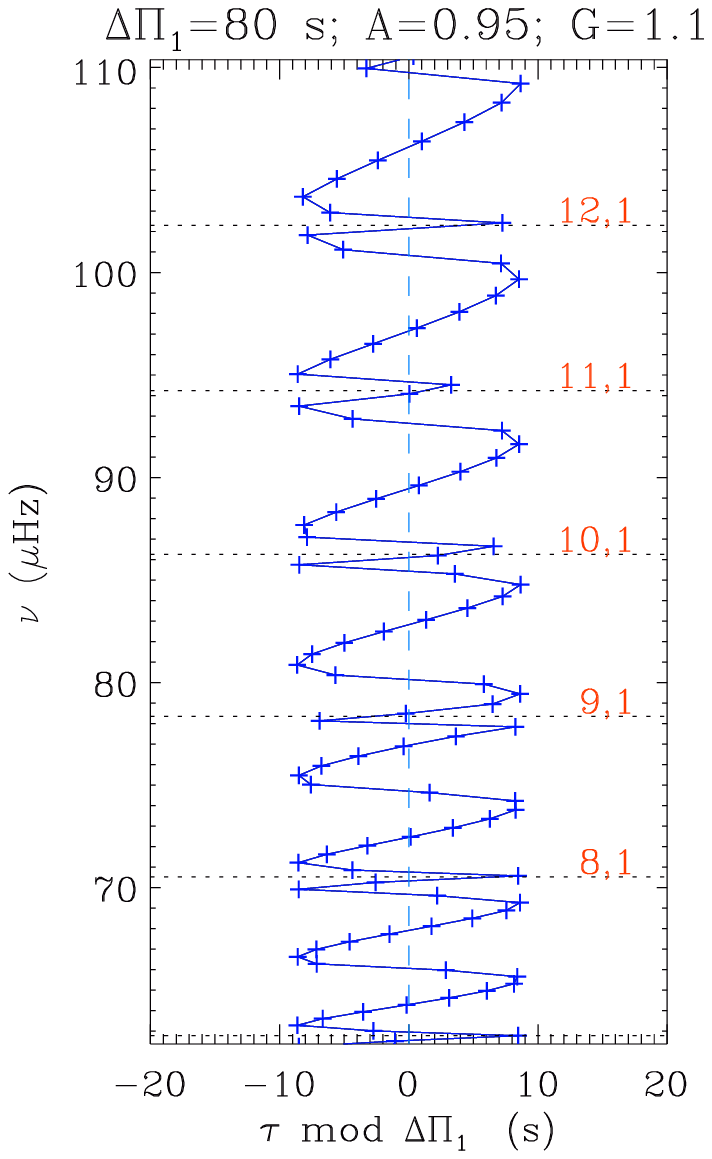}
\includegraphics[width=3.5cm]{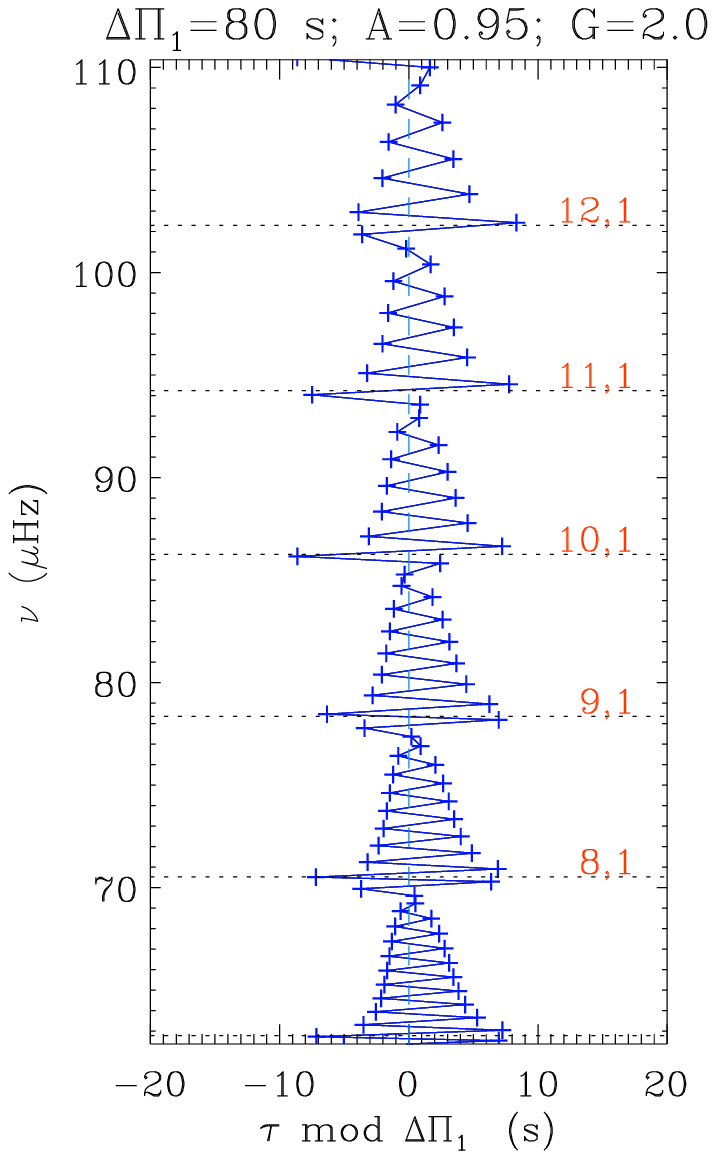}
\includegraphics[width=3.5cm]{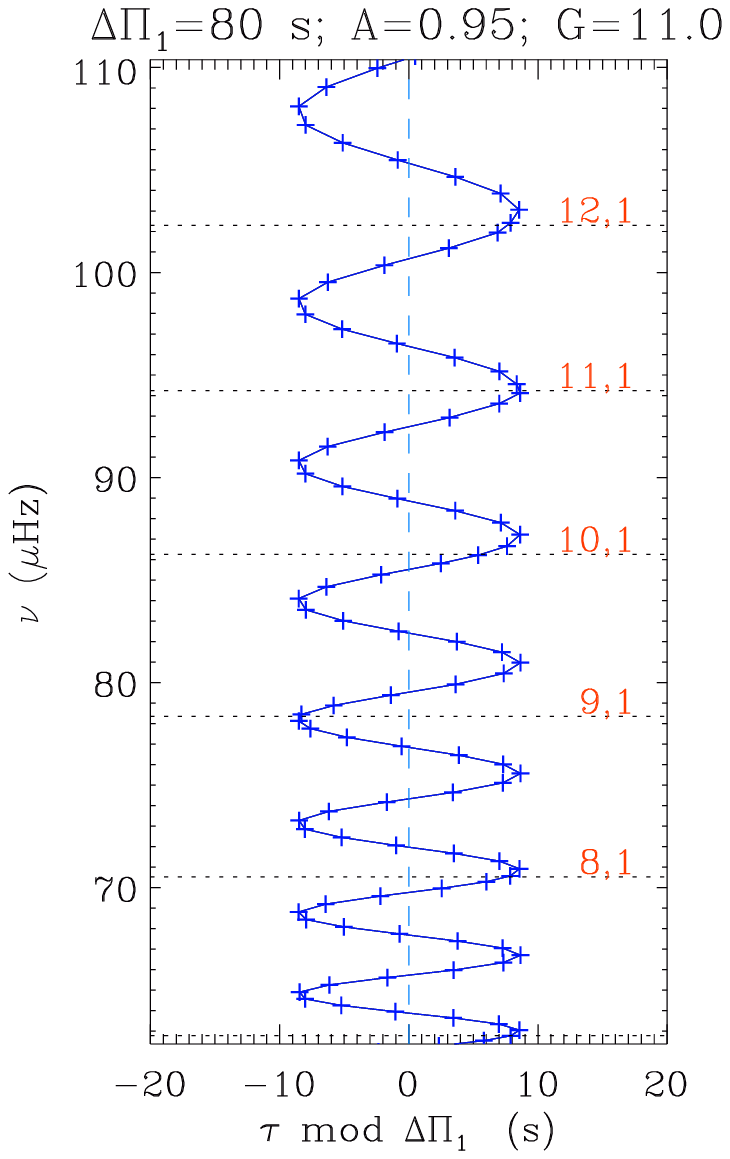}
\includegraphics[width=3.5cm]{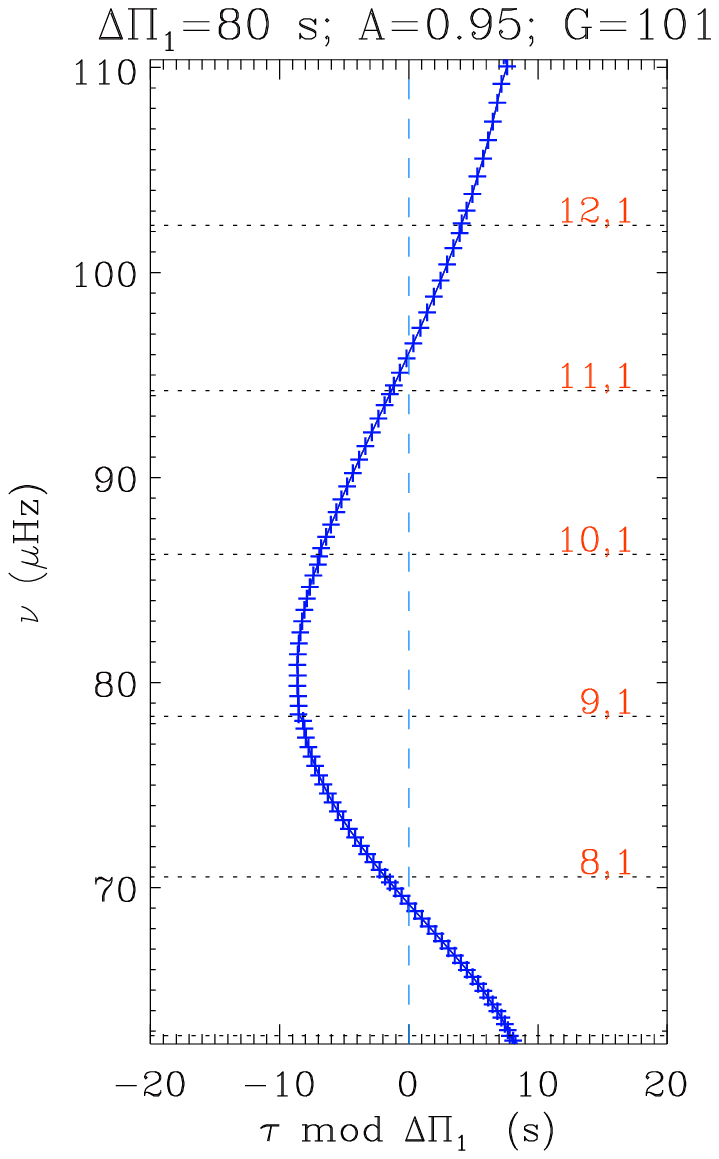}
\caption{Stretched period \'echelle diagrams of synthetic
mixed-mode spectra including buoyancy glitches. The period of the
buoyancy-glitch modulation is indicated for each plot: it
increases from the left to the right. All other parameters are the
same in all plots. Horizontal dotted lines indicate the location
of the pure pressure dipole modes.}\label{fig-glitchg}
\end{figure*}

When \BV\ glitches are present, the derivation of
Eq.~(\ref{eqt-asymp}) for obtaining Eq.~(\ref{eqt-zeta}) must take
the variation of $\glitchg$ into account. We suppose that the
buoyancy glitch can be treated as a perturbation, so that the
influence of the glitch in the phase $\thetap$ can be neglected.
Hence, Eq.~(\ref{eqt-deriv3}) is changed into
\begin{equation}\label{eqt-modul-change}
   {1\over \Tg} {\diff P\over \diff\nm} = \zeta - {\diff \glitchg\over\diff
   \nm} = \zetag.
\end{equation}
The correcting term on the right-side of this equation includes
the glitch contribution. We have checked that, except for huge
glitches, the glitch-perturbed correction $\zetag$ is very close
to the glitch-free function $\zeta$
(Fig.~\ref{fig-compare-zeta-glitch}). As a consequence, the
function $\zeta$ can be safely used in Eq.~(\ref{eqt-stretch})
instead of $\zetag$ to derive stretched periods. The stretched
periods calculated with $\zeta$ exhibit the glitch signature;
conversely, when $\zetag$ is used in Eq.~(\ref{eqt-stretch}), the
glitch signature is corrected. We also notice that the expression
used for the glitch is coherent with the modelling done by
\cite{2015ApJ...805..127C}.

\subsection{Deep and shallow buoyancy glitches}

We analyzed the cases where the function $\glitchg$ expressing the
buoyancy glitch varies sinusoidally and tested glitch periods
$\periodeglitch$ in a broad range, from 1.01 to 100. Even if the
sine form is a simplification, it helps investigating the glitch
signature since it carries the most important information, namely
the location of the glitch, which is related to the relative
period $\periodeglitch$. We denote the amplitude of the glitch as
in \cite{2008MNRAS.386.1487M}: the amplitude $A$ corresponds to a
relative jump $(1-A)$ in the \BV\ profile at the glitch location.

Figure \ref{fig-glitchg} shows that the \'echelle diagrams based
on the stretched period spacings make it possible to retrieve the
glitch profile. We note that the glitches with relative periods
$(\periodeglitch-1)$ and $1/(\periodeglitch-1)$ show similar
modulation, except near the p-m modes. Such a degeneracy in the
glitch information is well known: for symmetry reasons, deep and
shallow glitches cannot be distinguished. Here, the degeneracy is
broken by the p-m modes. We might distinguish three main
cases:\\
- For $\periodeglitch$ values close to unity, that is for glitches
located near the lower boundary of the \BV\ cavity, the period
bumping near the p-m modes induces a significant shift compared to
the period $\periodeglitch \Tg$ (Eq.~\ref{eqt-modul-glitchg}). As
a result, the large variation of the phase of the modulation
exceeds $2\pi$ so that it cannot be corrected. So, a rapid
variation of the stretched period near the p-m modes is the
signature of a shallow buoyancy glitch. The amplitude of the
spikes at the p-m modes can be used to investigate the amplitude
of the modulation of the glitch. According to
Fig.~\ref{fig-glitchg}, the spikes near the p-m modes have the
same amplitude as the global modulation. This result is robust; it
derives from a phase effect and is independent of the exact form
of the periodic function $\glitchg$.\\
- For periods $\periodeglitch$ larger than 2, the shift at the p-m
mode remains limited, so that the $\zeta$ contribution makes it
possible to retrieve the glitch function $\glitchg (1/\nu)$  in
the \'echelle diagram based on the stretched period. Large values
of $\periodeglitch$, corresponding to shallow glitches, show
long-period modulations. The absence of any accident near the p-m
mode in the \'echelle diagram is the signature of a shallow
glitch.\\
- The case $\periodeglitch =2$ is an intermediate case: the glitch
signature looks like a sawtooth profile; the varying amplitude of
the modulation is a moir\'e effect. Determining the mean value of
$\Tg$ may then be difficult. However, such cases were not
encountered in the large data set analyzed in
\cite{2014A&A...572L...5M}.

\begin{figure}
\centering
\includegraphics[width=5.8cm]{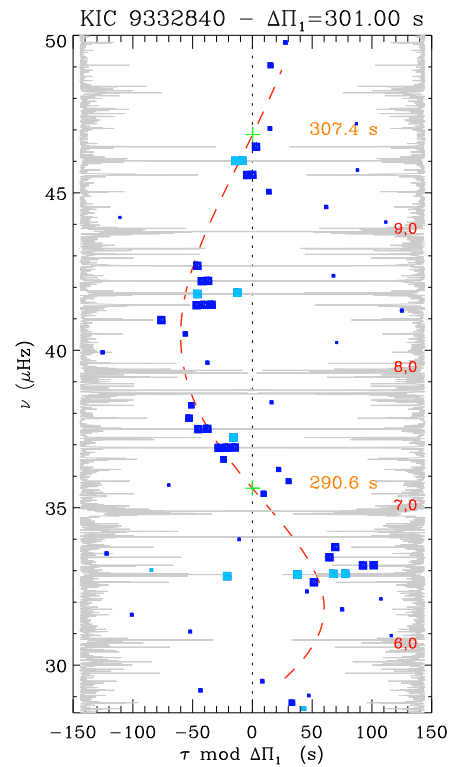}
\caption{Stretched period \'echelle diagram of KIC 9332840,
showing a large-period modulation (dashed red line) compatible
with a shallow buoyancy glitch. Peaks are identified as in
Figs.~\ref{fig-echelle} and \ref{fig-rotation}. The minimum and
maximum values of the varying period are indicated in orange.}
\label{fig-9332840}
\end{figure}

\subsection{An example of buoyancy glitch}

We consider the case of the red giant star KIC 9332840 observed by
\Kepler. This star has reached the red clump and shows a complex
mixed-mode pattern \citep{2012A&A...537A..30M}. Two values of the
period spacing were reported, depending on the frequency range:
298\,s in the low-frequency range and 306\,s in the high-frequency
range. This variation can be attributed to the modulation induced
by a buoyancy glitch (Fig.~\ref{fig-9332840}). An alternative
explanation should be a deviation to the first-order asymptotic
expansion. Observationally, the observed mixed modes have so high
gravity radial orders, about 80, that we do not favor this
hypothesis. Theoretical simulations by \cite{2015ApJ...805..127C}
seem to confirm this for RGB stars.

When the modulation is interpreted as a buoyancy glitch, the
stretched period diagram of this spectrum shows a signature with a
high $\periodeglitch$ value, in the range [40 -- 50]. This means
that the glitch has a shallow location and might be related to
some accident in the \BV\ profile near the outer boundary of the
dense radiative core. The amplitude $A$, about 0.08$\pm$0.02,
provides a measure of the glitch contrast in the \BV\ profile.

So, contrary to sound-speed glitches, buoyancy glitches may
significantly perturb the mixed-mode pattern. However, according
to previous observations \citep{2014A&A...572L...5M}, this only
occurs in a small minority of cases. A comprehensive observational
study of buoyancy glitches further requires a systematic analysis
of the red giant oscillation spectrum.

\section{Conclusion\label{conclusion}}

We have rewritten the asymptotic expansion of dipole mixed modes
in order to account for the most precise description of the pure
pressure and pure gravity contributions. We have shown that the
bumped period spacing and frequency rotational splitting follow
the same pattern constructed by the function $\zeta$
\citep{2015A&A...580A..96D}. This equality implies that the
information present in the rotation splittings and in the period
spacings is degenerated: both signatures are derived from the
ratio of the kinetic energy in the radiative core compared to the
total kinetic energy of the modes. Then, we have shown how the
function $\zeta$ can be used for stretching the mixed-mode pattern
and deriving the contribution of the pure gravity-mode pattern. In
fact, stretched periods mimic the pure gravity period spacing:
they are evenly spaced, and the mean spacing corresponds to the
asymptotic period spacing $\Tg$. As a result, each component of
the rotational multiplets can be identified and the asymptotic
period spacing can be measured even in presence of important
rotational splittings.

It follows that all steps of the red giant spectrum analysis can
be automated, since the complex mixed-mode forest is now as
regular as an artificially planted thicket. We have derived an
automated method for measuring $\Tg$ in a companion paper
\citep{vrard}; the automated measurement of the core rotation in
13\,000 red giant spectra observed by \Kepler\ and showing
solar-like oscillations is in progress.

Another output is the possible interpretation of the modulation of
the period spacing as buoyancy glitches. \'Echelle diagrams based
on the stretched periods are able to put them in evidence. This
reinforces the capability of mixed modes for probing the stellar
cores and inferring unique information on the physical conditions
in the nuclear-burning region.

%______________________________________________________________
%\begin{acknowledgements}
\begin{acknowledgements}
We acknowledge the entire Kepler team, whose efforts made these
results possible. We acknowledge financial support from the
Programme National de Physique Stellaire (CNRS/INSU) and from the
ANR program IDEE Interaction Des \'Etoiles et des Exoplan\`etes.
We thanks the referee, H. Shibahashi, for his constructive
comments.
\end{acknowledgements}

\bibliographystyle{aa} % style aa.bst
\bibliography{biblio_z}

\begin{appendix}
\section{Limiting cases\label{limite}}

\subsection{Gravity-dominated mixed modes}

The frequency $\nu$ of a gravity-dominated mixed mode (g-m mode)
is significantly offset compared to the pure pressure frequency
$\nup$, so that its phase $\thetap$ is close to $\pi/2$, modulo
$\pi$ and its cosine is small. Hence, the function $\zeta$ shows
large variation. For these g-m modes, $\tan\thetap$ is large, so
that $\tan\thetag$ is necessarily large too, according to the
asymptotic relation (Eq.~\ref{eqt-asymp}). Considering that both
cosine values are small, one can approximate them by the inverse
of their tangents. As a result, when taking the asymptotic
expansion into account, the value of $\zeta$ for g-m modes is
close to $ \zeta\ind{max}$ defined by
\begin{equation}\label{eqt-zeta-max}
    \zeta\ind{max}
    =
    \left[1+ q  {\nu^2 \Tg \over \Dnup} \right]^{-1}
    =
    \left[1+  {q \over \nmix} \right]^{-1} ,
\end{equation}
hence close to unity in most cases except on the lower part of the
red giant branch. As a by-product of this discussion, we note that
the g-m periods are close to $(\nm+1/2+\epsg) \Tg$. So, even if
g-m modes behave as gravity modes, their frequencies are shifted
due to an extra phase added by the coupling; this situation
resembles the $\pi$ phase added to the wavefront of an optical ray
crossing a focal point.

\subsection{Pressure-dominated mixed modes}

On the contrary, for pressure-dominated (p-m) mixed modes, the
phase $\thetap$ is close to 0: the function $\zeta$ shows reduced
variation between $\zeta\ind{min}$ and 1. The low value of
$\tan\thetap$ implies the low value of $\tan\thetag$;
$\cos^2\thetap$  and $\cos^2\thetag$ are both close to unity, so
that for p-m modes one finds $\zeta$ close to $\zeta\ind{min}$
defined by
\begin{equation}\label{eqt-zeta-min}
    \zeta\ind{min}
    =
    \left[1+ {1\over q} {\nu^2 \Tg \over \Dnup} \right]^{-1}
    =
    \left[1+  {1 \over q \nmix} \right]^{-1}
    .
\end{equation}
This term $\zeta\ind{min}$ is significantly different from unity;
as is well known, period spacings of p-m modes cannot be used to
derive the asymptotic period spacing
\citep[e.g.,][]{2011Natur.471..608B,2013ApJ...765L..41S}.
\end{appendix}

\end{document}